\documentclass{aa}  
\usepackage{overpic}
\usepackage{graphicx}
\usepackage{txfonts}
\usepackage{lipsum}
\usepackage{subcaption}         
\usepackage{lscape}             
\usepackage{placeins}           
\usepackage[colorlinks=true, linkcolor=blue, citecolor=blue, urlcolor=blue]{hyperref}
\begin{document}

   \title{Testing the black hole star model: A UV-to-optical spectral fitting of The Cliff}

\titlerunning{A UV-to-optical spectral fitting of The Cliff}
%
%
%

   \author{Rosa M. Mérida\inst{1}\corrauth{Rosa.MeridaGonzalez@smu.ca}
        \and Marcin Sawicki\inst{1} \and Gaia Gaspar\inst{1,2} \and Chris J. Willott\inst{3} \and Kartheik G. Iyer\inst{4}
        }

   \institute{  
            Institute for Computational Astrophysics and Department of Astronomy and Physics, Saint Mary's University, 923 Robie Street, Halifax, NS B3H 3C3, Canada \and
            Observatorio Astronómico de Córdoba, Universidad Nacional de Córdoba, Laprida 854, X5000, Córdoba, Argentina \and
            National Research Council of Canada, Herzberg Astronomy \& Astrophysics Research Centre, 5071 West Saanich Road, Victoria, BC, V9E 2E7, Canada \and
            Columbia Astrophysics Laboratory, Columbia University, 550 West 120th Street, New York, NY 10027, USA
            }

   \date{Received September 30, 20XX}

  \abstract
   {In the black hole star (BH$^*$) framework, the characteristic V-shaped spectral energy distribution of little red dots (LRDs) is produced by an accreting BH embedded in a dense neutral-gas envelope with a near-unity covering factor. This envelope reprocesses the radiation and emits as a $\sim$5,000~K blackbody, producing the optical continuum in LRDs. Meanwhile, the ultraviolet (UV) is powered by a low-mass, dust-free, metal-poor host. The BH$^*$ scenario is promising but has yet to undergo detailed testing; conducting a self-consistent UV-to-optical spectral-fitting analysis of LRDs would provide a robust assessment of the model. In this work, we tested the BH$^*$ scenario by fitting the full JWST/NIRSpec PRISM spectrum of ``The Cliff'' ($z_{\mathrm{spec}}=3.55$), an LRD that played a pivotal role in the development of the BH$^*$ model. We find that a \texttt{Bagpipes} fit that allows stellar, nebular, active galactic nucleus (AGN), and blackbody components naturally yields a BH$^*$-like solution for The Cliff, even with broad priors. Our method allows us to characterize its host despite it remaining unresolved in JWST imaging. From the continuum, we infer the host to be low-mass ($\log\,M_\star/M_\odot \,\sim7.7$), star-forming, metal-poor, and affected by non-negligible dust attenuation ($A_V\sim0.5$~mag) acting on both stellar and nebular components. Higher stellar masses (up to $\log\,M_\star/M_\odot \,\sim8.1$) and attenuations (up to $A_V\sim1$~mag) are obtained depending on the assumed dust attenuation law. Modest AGN UV leakage is consistently allowed---and sometimes preferred---by the code but remains weak and not robustly constrained, with both AGN+host and host-dominated UV scenarios yielding statistically equivalent fits. The star formation history of the host is relatively smooth, with the galaxy already assembling $\log\,M_\star/M_\odot\sim7$ about 200~Myr before $z_{\mathrm{spec}}=3.55$. The BH-to-$M_\star$ ratio consistently exceeds the values expected from BH--host scaling relations, especially at recent times. This tension may indicate either inaccurate estimates of the BH properties or non-coeval BH--host evolution in this LRD.}


   \maketitle
   \nolinenumbers

\section{Introduction}
\label{sec:intro}

Thanks to an increasing amount of multiband photometric and spectroscopic data spanning from X-ray to radio wavelengths, our view of the so-called little red dots (LRDs; \citealt{Labbe2023}, \citealt{Barro2024}, \citealt{Greene2024}, \citealt{Matthee2023}) has become much clearer in recent years. 
A broad set of observational properties has now been established for LRDs, including (i) their high-$z$ nature (i.e., their number density drops $1-2$ dex from $z\sim 4$ to $z \sim 2$; \citealt{Kocevski2025}, \citealt{Inayoshi2025b}, \citealt{Ma2025}, \citealt{Zhuang2025}); (ii) the absence of X‑ray, submillimeter, and radio emission (e.g., \citealt{Labbe2023}, \citealt{Ananna2024}, \citealt{Casey2024}, \citealt{Maiolino2024}, \citealt{Labbe2025}, \citealt{Perger2024}, \citealt{Setton2025}); (iii) their broad Balmer and Paschen lines; (iv) their characteristic V-shaped spectral energy distributions (SEDs); (v) their compactness; and (vi) their lack of variability (e.g., \citealt{Kokubo2025}, \citealt{Furtak2025}, \citealt{Zhang2024}). 

More recently, a new wave of discoveries has further complicated this landscape. These include (vii) the report of extremely large Balmer breaks, arguably not stellar, in some LRDs (e.g., \citealt{deGraaf2025}, \citealt{Naidu2025}); (viii) the presence of Balmer-line absorption in many LRDs (e.g., \citealt{Juodvzbalis2024}, \citealt{DEugenio2025}); (ix) the discovery of potential local analogs (e.g., \citealt{RLin2025, Lin2025}, \citealt{Ji2025b}); and (x) the detection of high-ionization emission lines in some LRDs (e.g., \citealt{Labbe2024environment}, \citealt{Akins2025}, \citealt{Torralba2026a}), particularly in the UV (\citealt{Labbe2024environment}, \citealt{Tang2025}, \citealt{Treiber2025}), as well as Fe\,II emission, characteristic of the broad line region (BLR) of classical active galactic nuclei (AGNs; e.g., \citealt{Labbe2024environment}, \citealt{Tripodi2025}, \citealt{Perez-Gonzalez2026}, \citealt{Torralba2026}).

However, despite the high level of detail in this increasingly complex picture, the physical nature, origin, and evolution of these sources remain unclear. Several studies have attempted to fit LRD SEDs using well-known codes and different combinations of stellar and AGN models, with the goal of placing these objects within the broader galaxy evolution framework (e.g., \citealt{Greene2024}, \citealt{Perez-Gonzalez2024},  \citealt{Merida2025}, \citealt{Tripodi2024}). Nevertheless, these attempts have failed to provide a unique prescription for LRDs, as the high stellar masses ($M_\star$) or dust attenuations ($A_V$) invoked, especially in those cases with extreme Balmer breaks, conflict with our current understanding of stellar physics, mass surface densities, and the lack of X-ray and infrared (IR) emission of LRDs.  

As a result, alternative solutions have been proposed, such as the black hole star (BH$^*$) model (e.g., \citealt{Inayoshi2024}, \citealt{Ji2025}, \citealt{Naidu2025}; see also \citealt{deGraaf2025} and \citealt{Rusakov2025} for supporting observational evidence), binary massive black holes \citep[BHs;][]{Inayoshi2025}, the super-Eddington unification model \citep{Madau2026}, direct collapse BHs (e.g., \citealt{Cenci2025}, \citealt{Baggen2026}, \citealt{Pacucci2026}), and globular clusters in formation \citep{Chisholm2026}. Many of these theories have common physical ingredients and could ultimately be pieces of a broader and better‑tailored framework for LRDs. However, they still require rigorous testing to determine whether they truly provide the solutions these sources demand.

One of the most popular theories in the current LRD paradigm is the BH$^*$ model. This idea envisions LRDs as accreting BHs whose emission is reddened by a dense envelope or cocoon of neutral gas ($n_H \sim 10^{9-11} \mathrm{cm}^{-3}$, $N_H \sim 10^{24-26} \mathrm{cm}^{-2}$) with a near-unity covering factor \citep{Torralba2026a}. Such a gas envelope would behave as a photosphere near the Hayashi limit, with temperatures of ${\sim 5,000 - 7,000}$~K (e.g., \citealt{Inayoshi2024}, \citealt{Kido2025}, \citealt{Inayoshic}). From a theoretical perspective, the origin of such an envelope is expected to be linked to super-Eddington accretion, whereas its disappearance could be driven by nuclear starbursts that expel the surrounding gas reservoir, effectively starving the envelope \citep{Inayoshic}. Observationally, the evolution of these sources can proceed in different ways, for example by growing a stellar outskirt \citep{Billand2025} or by losing the steep optical slope linked to the envelope \citep{Asada2026,Merida2025_stingray}, and a few LRDs might even survive to the present day in the local Universe. 

Key evidence in favor of the BH$^*$ model was the discovery of ``extreme'' LRDs such as The Cliff \citep{deGraaf2025}. This galaxy, located at $z_\mathrm{spec} = 3.55$, exhibits a distinctive feature that rules out a stellar origin for The Cliff and other similar LRDs. The Cliff displays a break close to the Balmer limit that mimics a stellar break and has an extremely high strength of 6.9. \citet{deGraaf2025} show that a stellar origin for this source is highly improbable, as its compact size and high $M_\star$ ($\sim10^{10.5}\,M_\odot$) imply that it should emit significant X-ray radiation. In \citet{deGraaf_review}, the optical continuum of this source was fitted using a modified blackbody (BB) component, which depends on a scale parameter, temperature, and a power law whose slope modifies the BB. They found a median best-fit temperature for the modified BB emission of The Cliff of $\sim3974$~K, which strongly supports the picture that LRDs are AGNs embedded in thermalized dense gas envelopes in approximate hydrostatic equilibrium.

However, the link between the rest-frame ultraviolet (UV) and optical emission of this now canonical LRD within the BH$^*$ model is not yet fully understood, nor is the origin of the LRD UV emission itself. Within the BH$^*$ interpretation, the UV emission of LRDs is primarily attributed to a low-mass, star-forming, metal-poor, and dust-free galaxy host (e.g., \citealt{Matthee2025}, \citealt{Pizzati2025}, \citealt{Lin2026} via clustering analysis; see also \citealt{Korber2026}, \citealt{Sun2026}; although see also \citealt{Barro2025b}, who report $M_\star>10^9\,M_\odot$ and $A_V\sim0.5$~mag for some LRDs). 

In many LRDs, the UV is spatially resolved, in line with the host hypothesis (e.g., \citealt{Chen2025}, \citealt{Rinaldi2024}, \citealt{Cloonan2026}). In fact, \citet{Cloonan2026} find that LRDs are typically more extended in the rest-frame UV than in the optical, with sources unresolved at both wavelengths being comparatively rare. This wavelength-dependent morphology is consistent with an extended stellar component contributing to the UV emission.
However, a purely stellar interpretation of the UV emission may be in tension with other spectroscopic signatures observed in some LRDs, including high-ionization UV lines and Fe\,II emission. This suggests that a variety of LRDs probably exist, with some being fully dominated by the host in the UV and others having some AGN contribution. The rest-frame UV emission of The Cliff is very compact, and there are no UV lines observed \citep{deGraaf2025}. Thus, the UV emission could have either a stellar or an AGN origin. 

So far, most observational LRD works that explore the constraints of the BH$^*$ model have focused either on a separate fit to the optical continuum emission or on illustrative models that combine star-forming galaxy spectra and a BB emission (e.g., \citealt{deGraaf2025}, \citealt{Ji2025}, \citealt{RLin2025}, \citealt{Taylor2025},  \citealt{Sun2026}, \citealt{Umeda2026}). These efforts have provided evidence in favor of the BH$^*$ model and shed light on the possible sources of the two regimes of the V-shaped SEDs. However, it is crucial to test whether the entire spectrum of LRDs can be well reproduced at once by the BH$^*$ model and if the retrieved physical properties---derived from the full posterior distributions rather than from visual agreement---are consistent with, on average, the LRD's attributes. 

A pioneering example of a UV-to-optical LRD spectral fitting was proposed in \citet{Barro2025b}, who fitted a sample of 118 LRDs (see also \citealt{Perez-Gonzalez2026}). They adopted a semiempirical model that combines a stellar and a BH$^*$ component, with the BH$^*$ template obtained via the fitting of a low-order polynomial to the emission line-free regions of The Cliff's spectrum, assuming no BH$^*$ contribution to the UV. The BH$^*$ template is modified per galaxy through the $A_V$ parameter only. 

\citet{Barro2025b} point out that the observed LRD diversity is primarily driven by variations in the  BH$^*$-to-galaxy luminosity ratio, together with an increasing obscuration of the BH$^*$ component. For The Cliff, they obtained a best-fitting stellar mass of log $M_\star/M_\odot = 8.67\pm0.63$ and a stellar extinction of $A_V = 0.45\pm0.18$~mag, which are consistent with a low- or intermediate-mass system subject to moderate levels of dust attenuation; the $A_V$ affecting the envelope is only $0.01\pm0.07$~mag.
However, that study did not consider an AGN component and assumed a BB nature for the optical continuum, with a fixed temperature across all of their LRDs. 

In this work, we used the spectral fitting code \texttt{Bagpipes} \citep{Carnall2018}, modified to include BB emission and a gas absorption component at the Balmer limit. Using PRISM data that probe the rest-frame UV and optical continua, we conducted a comprehensive spectral analysis of The Cliff, examining the ability of the BH$^*$ model to describe this LRD and analyzing its host properties.
Throughout this work, we assumed $\Omega_\mathrm{M,0}=0.3$, $\Omega_{\Lambda,0}=0.7$, and H$_0=70$ km s$^{-1}$ Mpc$^{-1}$. All $M_\star$ and star formation rate (SFR) estimates assume a \citet{Chabrier2003} initial mass function (IMF).

\section{Data and methodology}
\label{sec:method}

This work is based on \textit{James Webb} Space Telescope (JWST) observations of The Cliff obtained with the Near Infrared Spectrograph (NIRSpec; \citealt{Jakobsen2022}) in PRISM mode (R~$\sim$~100) using the micro-shutter assembly.
This galaxy corresponds to source $\#$154183 from the Red Unknowns: Bright Infrared Extragalactic Survey (RUBIES; PIs: A. de Graaff and G. Brammer; \citealt{deGraaf2025Rubies}). The data were obtained from version four of the DAWN JWST Archive (DJA; \citealt{Brammer2025}).

As mentioned in Sect.~\ref{sec:intro}, current SED fitting codes, based only on stellar and AGN emission, cannot properly fit LRDs. One of the most promising explanations of the nature of LRDs is based on the presence of a dense gas envelope that is capable of obscuring the emission from the central engine, which is, however, a feature that most standard SED fitting codes do not explicitly model at present. \citet{deGraaf_review} used a modified BB prescription to fit the optical continuum emission of a sample of LRDs (see also \citealt{Barro2025b}, \citealt{Perez-Gonzalez2026}, \citealt{Umeda2026}). This gas cocoon has also been modeled in several works using \texttt{Cloudy} (\citealt{Ferland2017}; see \citealt{Ji2025}, \citealt{RLin2025}, \citealt{Taylor2025} among many others). 

In this work, we attempted to fit the full SED of The Cliff using a customized version of \texttt{Bagpipes}. This \texttt{Python} software utilizes a Bayesian inference approach and allows the UV and optical emission from an AGN accretion disk to be fitted using a double power-law model. The code also implements nonparametric star formation histories (SFHs) using a series of piecewise constant functions in lookback time. We used the \texttt{continuity} mode, based on the nonparametric SFH formulation of \citet{Leja2019}, and all fits were performed using the Multinest sampling algorithm. A great advantage of \texttt{Bagpipes} is that it allows the user to include additional model components using the option \texttt{extra$\_$components} in the \texttt{fit$\_$dictionary} (see also \citealt{Taylor2025}). The sum of all the components we used results in a total of 19 free parameters. Through these components and a set of priors, we can test different key elements of the BH$^*$ model.

\begin{table*}[!htp]
\setlength{\tabcolsep}{1.6pt}  
\tiny
    \centering
    \caption{\texttt{Bagpipes} priors used in this work.}
    \begin{tabular}{c|c|c|c|c|c}
    \hline
         SFH&Nebular&Dust&AGN&BB&Absorption\\ \hline\hline
         \texttt{bin$\_$edges} = [0, 10, 30, 100, 300, 700, 900]&\texttt{log U} = [$-4, 0$]& $A_V$ = [0, 0.5] and&\texttt{alphalam} = [$-2.8, \,2$]&\texttt{T} = [1000, 7000]&\texttt{log sigma$\_$B0} = [$-18, -15$]\\
         \texttt{massformed} = [4, 8.5] and &&[0, 5]&\texttt{betalam} = [0.8, 1.6]&\texttt{log A} = [$-25, -16$]&\texttt{log N} = [19, 24]\\
         $[4, \,10]$&&&\texttt{hanorm} = $[0, \,5\times10^{-16}]$&&$\lambda_{break}$ = [3000, 4000]\\   
         \texttt{metallicity} = [0.001, 0.05] and&&&\texttt{f5100A} = [0, 3$\times10^{-16}$]&&\texttt{break$\_$width} = [10, 100]\\ 
         $[0.001, \,2.5]$& & & & & \\ \hline
    \end{tabular}
    \label{tab:priors}
    \tablefoot{\texttt{Massformed}, \texttt{metallicity}, and $A_V$ show two different ranges, corresponding to the BH$^*$+dust-free and dusty host configurations. The SFH is set to the \texttt{continuity} mode, and the bins are expressed in Myr. dSFR$_i$, which represents the change in log SFR between two consecutive SFH bins, is set to $[-10, 10]$, following a Student-t distribution. The \texttt{massformed} parameter is logarithmic and in $M_\odot$ units; the metallicity is logarithmic and in $Z_\odot$ units; $A_V$ is in magnitudes; \texttt{hanorm} has units of erg/s/cm$^2$ and \texttt{f5100A} is in erg/s/cm$^2/\AA$; $\lambda_{break}$ and \texttt{break$\_$width} are in units of $\AA$. The normalization factor of the BB, A, is expressed in erg/s/cm$^2/\AA$ and the temperature is in K. In the absorption component, the \texttt{sigma$\_$B0} $\times$ \texttt{N} product is dimensionless. In total, we work with 19 free parameters.}
\end{table*}

For the priors, we adopted two main configurations: one called BH$^*$+dust-free host and another called BH$^*$+dusty host. These configurations are identical in all parameters except those that affect the stars, stellar metallicity, and dust of the stellar component. 
The BH$^*$+dust-free host configuration aims to align with the expectation of the BH$^*$ model for the host of LRDs, indicating a metal-poor, dust-free, low-mass galaxy. The BH$^*$+dusty host configuration explores a broader set of priors for the stellar component. If the former accurately represents nature, then the two configurations should converge to the same low-mass, metal-poor, and dust-free solution. 

We ran \texttt{Bagpipes} using these two configurations, each under two scenarios: (i) including the main optical-to-near IR emission lines, and (ii) masking them out. The emission lines were masked by assigning large uncertainties to both the lines and the surrounding continuum. This approach allows us to assess whether any discrepancies between the BH$^*$+dust-free and dusty host configurations arise from constraints imposed by the continuum or by the emission lines.
The redshift was set to $z_{\mathrm{spec}} = 3.546$, as provided by the DJA and also specified in \citealt{deGraaf2025}. All the priors are flat, and a summary of all the prior configurations is provided in Table~\ref{tab:priors}.

\paragraph{\textbf{Blackbody emission}.}

We can model the emission from a BB analytically using Planck's law. In that way, we can introduce this component with just two parameters: temperature and a normalization factor, which accounts for the contribution of this component to the total solution. We set the allowed temperature range to T = [1000, 7000] K and the normalization range to log A$_{\mathrm{BB}}$~=~[$-25, -16$], with A$_{\mathrm{BB}}$ expressed in units of erg/s/cm$^2/\AA$.

\paragraph{\textbf{Absorption}.}

In The Cliff, the break located at the Balmer limit is very extreme (Balmer strength of 6.9), as a result of gas absorption. This is translated into a sharp drop in the optical continuum emission, which in our formulation should manifest as an equally sharp decline in the BB component at the Balmer limit. However, in the nominal Plank function the tail of this component does not fall off rapidly enough to reproduce such strong breaks. As a result, the BB tail extends into the UV unless it is suppressed explicitly.

To reproduce potential sharp spectral discontinuities observed around the Balmer limit, we introduced a phenomenological, data-driven absorption model analogous to the treatment commonly adopted for the Lyman break:
\begin{equation}
\begin{aligned}
    F_{final,\, BB}(\lambda) = B(\lambda) \,T(\lambda); \\
    T(\lambda) = e^{-\tau_B(\lambda)};\\
    \tau_{B} = N\,\sigma_B(\lambda);\\
    \sigma_{B} = \sigma_{B_0} \left(\frac{\lambda}{\lambda_{break}}\right)^3\,\sigma(\lambda);\\
    \sigma(\lambda) = \frac{1}{ 1+e^ { \left(\frac{\lambda-\lambda_{break}}{w}\right)} }.
\end{aligned}
\end{equation}
In the above formulation, the intrinsic BB spectrum, $B(\lambda)$, is attenuated by a transmission function $T(\lambda)$, where $\tau_B$ represents an effective optical depth associated with the absorption of the Balmer continuum. We parametrized this optical depth as a function of $N$, an effective column density, and $\sigma_B(\lambda)$, a wavelength-dependent cross section. $N$ is a column density, but it does not correspond directly to a physical $N_H$; rather, it acts as a parameter that controls the strength of the break. Motivated by the bound-free absorption of hydrogen, the cross section is assumed to scale as $\sigma_B(\lambda) \propto (\lambda / \lambda_{\rm break})^3$ below the Balmer edge. To avoid a nonphysical, overly sharp discontinuity, we introduced a smooth transition function modeled as a logistic function, $\sigma(\lambda)$, with characteristic width $w$, which regulates the onset of absorption across the break. In the fit, $\lambda_{\rm break}$ is allowed to vary around the vicinity of the Balmer limit. This absorption feature ($T(\lambda$)) and the BB spectrum are combined to produce the emission of the envelope, represented as a single curve in all our figures (i.e., $F_{final,BB}(\lambda)$). 

In Appendix~\ref{app2} we present a fit to The Cliff without this absorption feature to demonstrate its effect. Without the gas absorption, the code attributes all the continuum emission to the BB, including the UV continuum. As a result, \texttt{Bagpipes} effectively eliminates the galaxy component in the fit. Without the absorption implemented in the model, LRDs such as The Cliff could be interpreted as envelope-dominated systems, with absent or negligible hosts. However, this setting does not yield a robust fit, as highlighted by the residuals and the reduced $\chi^2$ values, emphasizing the need for the inclusion of the absorption component.

Nevertheless, we note that, as the Balmer opacity decreases toward shorter wavelengths, a sufficiently hot BB continuum can reemerge in the UV and contribute to the observed UV emission (see \citealt{Zhang2025}, \citealt{Sun2026}). Thus, a single-BB approximation may not fully capture the potential contribution of the gas envelope to the UV emission of the general LRD population.

\paragraph{\textbf{Stars + nebular emission + dust attenuation}.}

The stellar metallicity prior was set to span $\log Z/Z_\odot$ = [0.001, 0.05] in the BH$^*$+dust-free host configuration, forcing a low-$Z$ solution, and extended to [0.001, 2.5] in the BH$^*$+dusty host configuration. In terms of stellar mass, LRD hosts are expected to be low-mass sources (see Sect.~\ref{sec:intro}), with ${M_\star\sim10^{7-8}\,M_\odot}$. In the BH$^*$+dust-free host configuration, log~$M_\star/M_\odot$ was set to span $[4.0,\,8.5]$. In the BH$^*$+dusty host mode, this prior was broadened to $[4.0,\,10.0]$. 

Nebular emission was incorporated through the ionization parameter, log U, whose prior was set to span [$-4$, 0]. We selected a Small Magellanic Cloud (SMC) extinction curve \citep{Gordon2003}, which acts only on the nebular and stellar emission, with $A_V$ values ranging over $[0, \,0.5]$~mag and $[0, \,5]$~mag in the BH$^*$+dust-free and dusty host configurations, respectively. We explored selecting a \citet{Calzetti2000} dust attenuation law instead in Sect~\ref{sec:discussion_dust}.

\paragraph{\textbf{AGN}.}

The AGN continuum emission was modeled as a broken power law characterized by three parameters: two spectral slopes ($\alpha_\lambda,\,\beta_\lambda$) and the continuum flux at the break point (5100~$\AA$; \texttt{f5100A}). The AGN line emission was included as H$\beta$ and H$\alpha$ lines. The intensities were left free (\texttt{hanorm}), while the line widths were fixed through the H$\alpha$ line width as measured by us from the spectrum. 

According to the BH$^*$ model, the dense gas envelope reprocesses the AGN emission, thereby eliminating any potential AGN contribution to the UV and optical continua. Although the origin and nature of the broad Balmer lines is still a matter of debate---propositions include electron scattering through nuclear dense plasma \citep{Rusakov2025}, BLR stratification \citep{Scholtz2026}, hard, strongly ionizing intrinsic SEDs and an anisotropic radiation field \citep{Madau2026}, and young massive stars surrounding the envelope \citep{Asada2026}---in this work we assumed that they originate in the BLR, as is the case for Type I AGNs. We acknowledge that this is a caveat in our test to probe the BH$^*$ model and that additional mechanisms for the broad Balmer emission should be explored in future works. 

However, the \texttt{Bagpipes} AGN model does not include hydrogen lines beyond H$\alpha$ and H$\beta$. As a result, any additional line emission is necessarily attributed to the host galaxy in the fit. Furthermore, \texttt{Bagpipes} uses Gaussian profiles for the broad line components and does not capture more complex line profiles, such as those observed in the Balmer lines of The Cliff \citep{deGraaf2025}. Repeating the fits with the lines masked, considering only the flux from the continuum (see Sect.~\ref{sec:results}), ensures that the emission lines---potentially contaminated by the AGN---are not key drivers in our modeling.

The contribution of the AGN continuum was set through the normalization parameter \texttt{f5100A} whose prior spans [0, $3\times10^{-16}$] erg/s/cm$^2/\AA$, with \texttt{f5100A} being a key parameter in our modeling. A low AGN normalization favors a stellar origin for the UV continuum within the model. In this configuration, \texttt{Bagpipes} is free to explore both AGN-dominated and AGN-suppressed solutions. A key point here is that a preference for low \texttt{f5100A} values would therefore provide independent support for the BH$^*$ scenario as it would emerge from the data rather than being enforced by the adopted priors.

The slopes priors were set to span $\alpha_\lambda$ = [$-2.8,\, 2$] and ${\beta_\lambda\,=\,[0.8,\, 1.6]}$. These values enable the AGN to produce steep UV slopes (i.e., a quasar-like UV spectrum with an unobscured accretion disk) while maintaining a weak optical continuum. Such steep UV slopes are not allowed within the BH$^*$ framework; however, we want to test whether the code can fully reproduce the UV with stars, or else it needs to invoke an AGN model with a steep UV slope. 

Any UV leakage from the AGN would challenge the current BH$^*$ model assumption of a fully suppressed AGN component. This possibility, however, cannot be ruled out given the presence of high-ionization UV lines in some LRDs (see Sect.~\ref{sec:intro}), which suggests that ionizing radiation can escape along certain lines of sight (e.g., due to a clumpy envelope or a nonspherical configuration; see \citealt{Hviding2026}, \citealt{Sok2026}).
However, with this test and these priors, we are implicitly assuming that if the gas envelope does not suppress AGN-powered lines, it will also fail to suppress part of the AGN UV continuum; this assumption may not hold if these lines are emitted from less buried regions (see \citealt{Torralba2026}). 

Moreover, although high-ionization UV emission is typically associated with AGN activity, this is not the only way these lines could be powered. The dense environments in which such BH$^*$ structures may form could also favor the formation of very massive or supermassive stars (\citealt{Chisholm2026}, \citealt{Kokorev2026}, \citealt{Nandal2026}, \citealt{Rantala2026}, \citealt{Williams2026}), which may contribute to the production of high-ionization UV lines. Their powerful stellar winds and outflows could additionally produce broadened high-ionization features.
However, testing these more complex and alternative scenarios is beyond the scope of this work.

\section{Results}
\label{sec:results}

\begin{figure*}[h!]
    \centering
    \sidecaption
        \parbox{0.70\textwidth}{
    \includegraphics[width=\linewidth]{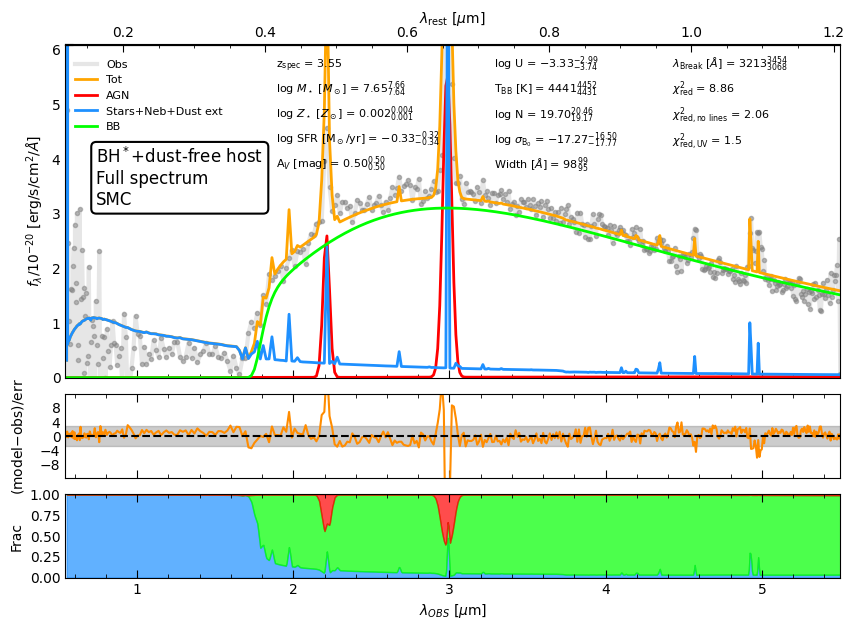}\\[-2mm]
    \includegraphics[width=\linewidth]{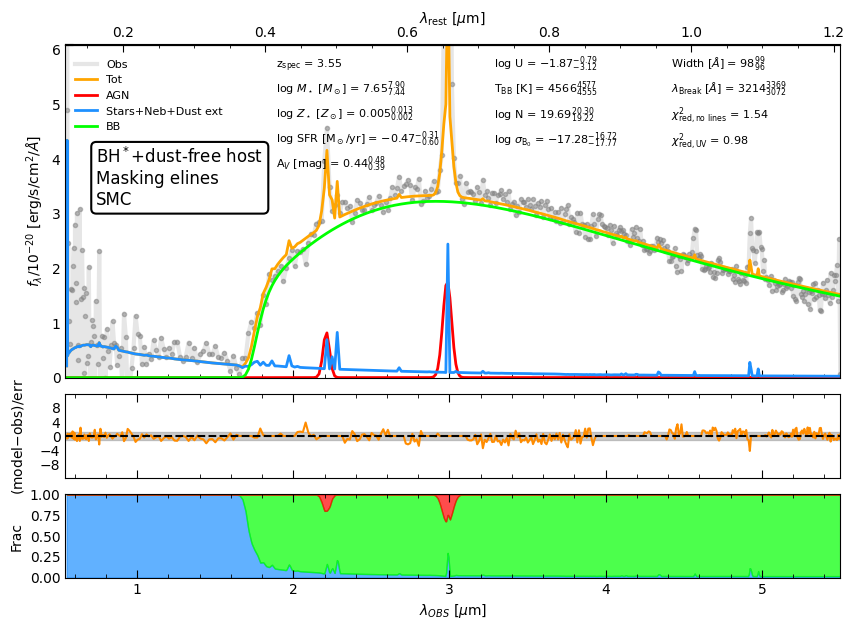}
}
    \caption{Best-fitting models from \texttt{Bagpipes} using the BH$^*$+dust-free host configuration for The Cliff. The top panel shows the fit based on the full spectrum, whereas the bottom panel considers only the flux from the continuum. The observed spectrum is shown in gray in both panels. We include the total best-fitting model (orange), the AGN component (red), the stellar + nebular component, attenuated by dust (blue), and the composite of the BB and Balmer absorption (green). The stellar + nebular model matches the total best-fitting model in the UV. We include three reduced $\chi^2$ values: based on the full spectrum, based only on the continuum, and measured only in the UV. The residuals of the fit are depicted underneath each SED plot. The shaded gray region represents the standard deviation. The fractional contribution of each component to the total best-fitting model is displayed in the bottom rows, following the same color code as above. These fits indicate that The Cliff requires a moderate level of dust attenuation.}
    \label{fig:cliff_pure}
\end{figure*}

\begin{figure*}[!htp]
\centering
\sidecaption
        \parbox{0.70\textwidth}{
    \includegraphics[width=\linewidth]{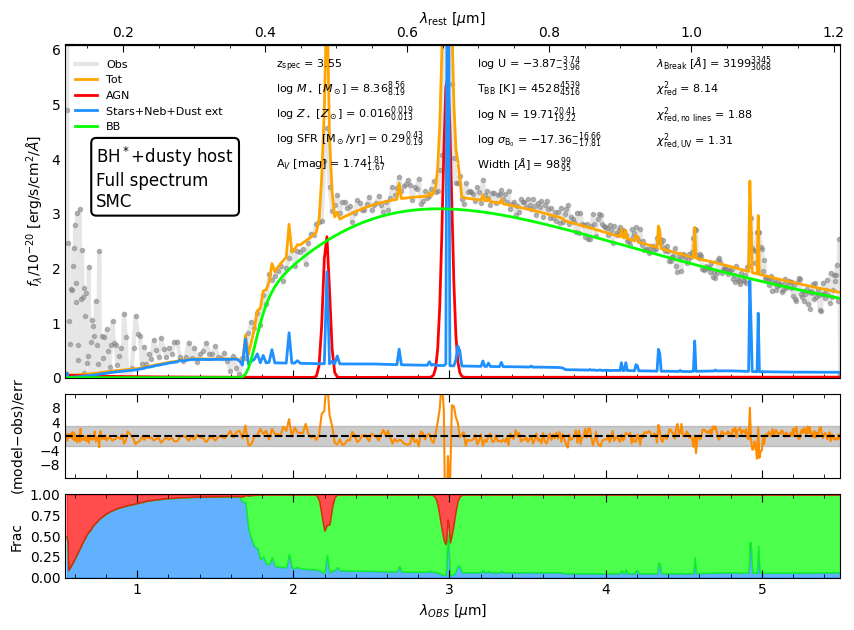}\\[-2mm]
    \includegraphics[width=\linewidth]{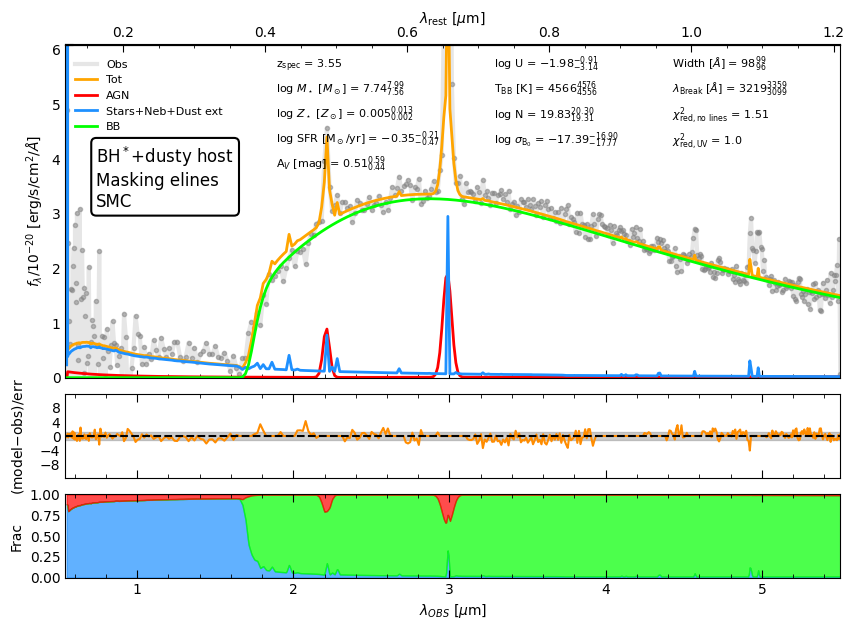}
}
    \caption{Same as Fig. \ref{fig:cliff_pure} but for the BH$^*$+dusty host configuration. Results based on the continuum match with those obtained with the BH$^*$+dust-free configuration, but higher $A_V$ and $M_\star$ values are retrieved when the full spectrum is considered.}
    \label{fig:cliff_dusty}
\end{figure*}

The \texttt{Bagpipes} best-fitting models for The Cliff according to both configurations, BH$^*$+dust-free and dusty hosts, are depicted in Figs.~\ref{fig:cliff_pure} and \ref{fig:cliff_dusty}, respectively. In each figure, we show the fits based only on the continuum and on the full spectrum, including the lines.
A list of the values for all parameters can be found in Table~\ref{tab:properties}. 

A key result is that across all fits and configurations, and despite wide priors, \texttt{Bagpipes} consistently converges toward a low AGN normalization, effectively suppressing the AGN continuum and assigning most of the UV emission to stars and most of the optical emission to the BB component. This is exactly in line with the expectations of the BH$^*$ model. It is very compelling evidence for the validity of the BH$^*$ model for this object, as this result is not based on any prior assumptions.

The fit also informs us about the properties of the host galaxy.
Looking at Fig.~\ref{fig:cliff_pure}, which shows the results for the BH$^*$+dust-free host configuration, we find that The Cliff is described as a ${\sim10^{7.7}\,M_\odot}$ galaxy with $A_V$ close to the upper limit of the prior (0.5~mag). Some level of dust attenuation is therefore required by the fit, effectively ruling out an $A_V=0$~mag scenario for this LRD.

The log SFR$_{100}$ values (which correspond to the SFR measured over the past 100 Myr) locate the host within the main sequence (MS) scatter of the \citet{Merida2025_ms} MS relation at the low-mass end at this redshift. Meanwhile, the BB temperature varies between 4,441 and 4,566~K, depending on whether emission lines are included, while the parameters defining the gas absorption are similar in the two cases.

The main difference between the two panels in Fig.~\ref{fig:cliff_pure} lies in the UV continuum fit. When fitting the full spectrum, including the emission lines (top panel), the model tends to overestimate the UV continuum, particularly at rest-frame $\lambda>0.2\,\mu$m, as reflected in the residuals and a reduced $\chi^2$ value as measured in the UV ($\chi^2_{\mathrm{red},\, \mathrm{UV}}$) of 1.5. In contrast, when masking the emission lines and fitting only the continuum (bottom panel), the model slightly underestimates the UV flux but achieves a better overall match with the data, especially in the UV, with a $\chi^2_{\mathrm{red,\,UV}}$ = 0.98.

\begin{table*}[htp]
\small
    \centering
    \renewcommand{\arraystretch}{1.4}
    \setlength{\tabcolsep}{3pt}  
    \caption{Properties of The Cliff as measured in this work. }
    \begin{tabular}{|c|c|c|c|c|c|c|}
    \hline
       &\multicolumn{2}{|c|}{BH$^*$+dust-free host} & \multicolumn{4}{c|}{BH$^*$+dusty host} \\ \hline
       &\multicolumn{2}{|c|}{SMC}&\multicolumn{2}{c|}{SMC}&\multicolumn{2}{c|}{Calzetti}\\\hline
       &Full spectrum&Continuum&Full spectrum&Continuum&Full spectrum&Continuum\\
       \hline\hline
         log $M_\star/M_\odot$&$7.65_{7.64}^{7.66}$&$7.65_{7.44}^{7.90}$&$8.36_{8.19}^{8.56}$&$7.74_{7.56}^{7.99}$&$8.26_{8.18}^{8.41}$&$8.12_{7.95}^{8.30}$\\ \hline
         log $Z/Z_\odot$&0.002$_{0.001}^{0.004}$&0.005$_{0.002}^{0.013}$&0.016$_{0.013}^{0.019}$&0.005$_{0.002}^{0.013}$&0.015$_{0.012}^{0.018}$&0.013$_{0.005}^{0.020}$\\ \hline 
         log U &$-3.33_{-3.74}^{-2.99}$&$-1.87_{-3.12}^{-0.79}$&$-3.87_{-3.96}^{-3.74}$&$-1.98_{-3.14}^{-0.91}$&$-3.88_{-3.96}^{-3.76}$&$-1.95_{-3.09}^{-0.96}$\\ \hline
         $A_V$ [mag]&$0.50_{0.50}^{0.50}$&$0.44_{0.39}^{0.48}$&$1.74_{1.67}^{1.81}$&$0.51_{0.44}^{0.59}$&$1.78_{1.72}^{1.84}$&$1.04_{0.94}^{1.14}$\\\hline
        log SFR$_{100}$ [$M_\odot$/yr]&$-0.33_{-0.34}^{-0.32}$&$-0.47_{-0.60}^{-0.31}$&$0.29_{0.19}^{0.43}$&$-0.35_{-0.47}^{-0.21}$&$0.24_{0.20}^{0.35}$&$-0.04_{-0.17}^{0.08}$\\\hline
        $\alpha_\lambda$&$0.18_{-1.37}^{1.47}$&$-0.38_{-1.76}^{1.09}$&$-2.74_{-2.78}^{-2.67}$&$-2.37_{-2.63}^{-2.08}$&$-0.70_{-2.08}^{0.77}$&$-0.93_{-2.00}^{0.35}$\\ \hline
        $\beta_\lambda$&$1.14_{0.94}^{1.41}$&$1.13_{0.91}^{1.42}$&$0.85_{0.81}^{0.93}$&$1.08_{0.91}^{1.29}$&$1.15_{0.94}^{1.38}$&$1.15_{0.95}^{1.37}$\\ \hline
         \texttt{hanorm} [erg/s/cm$^2$]&3.98$_{3.89}^{3.98}\times10^{-17}$ &8.32$_{3.39}^{13.49}\times10^{-18}$&3.89$_{3.80}^{3.89}\times10^{-17}$&8.51$_{4.47}^{13.80}\times10^{-18}$&3.89$_{3.80}^{3.89}\times10^{-17}$&9.12$_{4.47}^{14.13}\times10^{-18}$\\ \hline
         \texttt{f5100} [erg/s/cm$^2/\AA$]&2.88$_{0.78}^{7.24}\times10^{-23}$ &5.01$_{1.74}^{11.22}\times10^{-23}$&1.00$_{0.81}^{1.20}\times10^{-21}$&7.41$_{2.34}^{17.38}\times10^{-23}$&5.13$_{1.55}^{10.96}\times10^{-23}$&5.89$_{1.95}^{12.59}\times10^{-23}$\\ \hline
         T$_{\rm{BB}}$ [K]&$4441_{4431}^{4452}$&$4566_{4555}^{4577}$&$4528_{4516}^{4539}$&$4566_{4556}^{4576}$&$4499_{4489}^{4509}$&$4573_{4564}^{4582}$\\\hline
         A$_{\rm{BB}}$ [erg/s/cm$^2/\AA$]&$1.41_{1.41}^{1.45}\times10^{-19}$& $1.48_{1.48}^{1.48}\times10^{-19}$&$1.41_{1.41}^{1.41}\times10^{-19}$&$1.48_{1.48}^{1.48}\times10^{-19}$&$1.41_{1.41}^{1.41}\times10^{-19}$&$1.48_{1.48}^{1.48}\times10^{-19}$\\\hline
         log N&$19.70_{19.17}^{20.46}$&$19.69_{19.22}^{20.30}$&$19.71_{19.22}^{20.41}$&$19.83_{19.31}^{20.30}$&$19.72_{19.26}^{20.42}$&$19.62_{19.24}^{20.21}$\\\hline
         log $\sigma_{B_0}$&$-17.27_{-17.77}^{-16.50}$ &$-17.28_{-17.77}^{-16.72}$&$-17.36_{-17.81}^{-16.66}$&$-17.39_{-17.77}^{-16.90}$&$-17.33_{-17.71}^{-16.77}$&$-17.34_{-17.74}^{-16.79}$\\\hline
         Width [$\AA$] &98$_{95}^{99}$&98$_{96}^{99}$&98$_{95}^{99}$& 98$_{96}^{99}$&98$_{95}^{99}$&98$_{96}^{99}$\\\hline
         $\lambda_{break}$ [$\AA$]& 3213$_{3077}^{3454}$&3214$_{3072}^{3369}$&3199$_{3068}^{3345}$& 3219$_{3099}^{3359}$&3210$_{3078}^{3377}$& 3213$_{3088}^{3405}$\\\hline
         $\chi^2_{\mathrm{red}}$&8.86&   &8.14&   &8.25&   \\\hline
         $\chi^2_{\mathrm{red,\,no\;lines}}$&2.06&1.54&1.88&1.51&1.84&1.46\\\hline
         $\chi^2_{\mathrm{red,\,UV}}$&1.50&0.98&1.31&1.00&0.89&0.96\\
         \hline
    \end{tabular}
    
    \label{tab:properties}
    \tablefoot{The values correspond to the median and 16th and 84th percentiles. We distinguish between the results obtained using an SMC (Sect.~\ref{sec:results}) and a Calzetti (Sect.~\ref{sec:discussion_dust}) dust attenuation laws, and we further separate the solutions derived from continuum‑only fits from those based on the full spectrum. SFR$_{100}$ corresponds to the SFR measured over the past 100 Myr. We include the reduced $\chi^2$ values associated with the best-fitting models. $\chi^2_{\mathrm{red,\,no\;lines}}$ values were computed considering only the emission from the continuum, and $\chi^2_{\mathrm{red,\,UV}}$ refers specifically to the UV continuum.}
\end{table*}

The BH$^*$+dusty host configuration, shown in Fig.~\ref{fig:cliff_dusty}, matches the BH$^*$+dust-free fit only if the emission lines are not taken into account. If the full spectrum is used, the code provides even more dust attenuation ($A_V\sim1.8$~mag, well outside the limits of the BH$^*$+dust-free host configuration prior), higher $M_\star$ (${\sim10^{8.4}\,M_\odot}$), and slightly higher metallicity (log $Z/Z_\odot\sim0.02$). We also obtain a higher SFR$_{100}$ value, close to the limit defined by the MS scatter. 

The BB temperature ranges from 4,528 to 4,566~K, depending on the inclusion of the emission lines. The parameters defining the absorption are broadly consistent across configurations. The reduced $\chi^2$ values are slightly better in this configuration than in the BH$^*$+dust-free host case. However, the fit in the UV is comparatively worse when the full spectrum is fitted, being too suppressed at shorter wavelengths.

In summary, based on the analysis of the continuum, we found that The Cliff host is likely a low-mass ($10^{7.7}\,M_\odot$), star-forming galaxy, characterized by a low metallicity (log $Z/Z_\odot$ = 0.005), and a moderate level of dust $A_V\sim0.5$~mag, with a gas envelope temperature of $\sim4,566$ K. If emission lines are considered, the dust attenuation and the $M_\star$ increase ($A_V\sim1.8$~mag and $M_\star\sim10^{8.4}\,M_\odot$), as well as the SFR$_{100}$.

We next compared our results with previous analyses of The Cliff.
\citet{deGraaf_review} fitted a modified BB component to the optical part of The Cliff's PRISM spectrum (see Sect.~\ref{sec:intro}) and found a median BB temperature, along with the associated 16th and 84th percentiles, of $3,974_{3,919}^{4,033}$ K. The temperatures obtained using both their method and ours are consistent with those expected for a photosphere near the Hayashi limit \citep{Kido2025}. As for the other physical parameters derived for The Cliff in this work, \citet{deGraaf_review} did not explore stellar or AGN properties in an SED-fitting context.

\citet{Barro2025b} reported best-fitting values for the $M_\star$ and attenuation of The Cliff using the SMC dust law. They found $\log(M_\star/M_\odot) = 8.67 \pm 0.63$ and $A_V = 0.45 \pm 0.18$ mag. Their $A_V$ estimate is consistent with our results, but their $M_\star$ is higher than our values, even within uncertainties. A comparable $M_\star$ is only recovered by our fits when adopting the BH$^*$+dusty host configuration and fitting the full spectrum, without masking the lines. However, this solution requires a much higher attenuation, $A_V\sim1.7$ mag, than that found by \citet{Barro2025b}. In addition, under these premises, \texttt{Bagpipes} assigns all emission line features, except the broad H$\alpha$ and H$\beta$ components, to stellar processes. This effectively maximizes the contribution of the stellar component to the fit. Consequently, this scenario should be regarded as an upper limit on the $M_\star$, suggesting that the true $M_\star$ is likely below $10^{8.4}\,M_\odot$.

Furthermore, as an independent consistency check, we compared our inferred stellar masses with the ${M_{\rm UV}-M_\star}$ relation at $z\sim3.8$ from \cite{Qiu2018}. Assuming that the observed UV emission is entirely stellar, we measure ${M_{\rm UV}=-17.3}$ for The Cliff, which corresponds to ${M_\star=10^{7.7}\,M_\odot}$, in excellent agreement with our continuum-based fits.

\section{Discussion}
\label{sec:discussion}

\subsection{The dusty nature of The Cliff's host}
\label{sec:discussion_dust}

As a result of our fits, we reject a pure dust-free scenario for this LRD. The required level of dust attenuation depends heavily on the emission lines, but it is never $A_V\sim0$~mag. Without any line emission constraints, the code requires $A_V \sim0.5$~mag. This could imply varying levels of dust acting on the stellar and nebular components depending on the stellar contribution to the lines. The maximum value would be an $A_V\sim1.8$ mag, corresponding to the maximum possible contribution of the host to the emission lines.

These results are based on assuming an SMC extinction law, commonly used in LRD studies (e.g., \citealt{Akins2024}, \citealt{Barro2025b}, \citealt{Ji2025}, \citealt{Taylor2025}, \citealt{Jones2026}). In this subsection we explore the effect of this assumption on the dust budget of The Cliff.
For this goal, we repeated our analysis choosing a \citet{Calzetti2000} dust attenuation law, also used in a variety of LRD studies (e.g., \citealt{Chen2025}, \citealt{deGraaf2025}, \citealt{Kocevski2025}, \citealt{Rinaldi2024}, \citealt{Umeda2026}). \citet{Tripodi2024} found this law to be consistent with an LRD at $z\sim8.6$ using a flexible analytical attenuation model that was broadly compatible with a Calzetti-like curve, although slightly shallower in the rest-frame UV.

\begin{figure*}[!htp]
    \centering
    \sidecaption
        \parbox{0.70\textwidth}{
    \includegraphics[width=\linewidth]{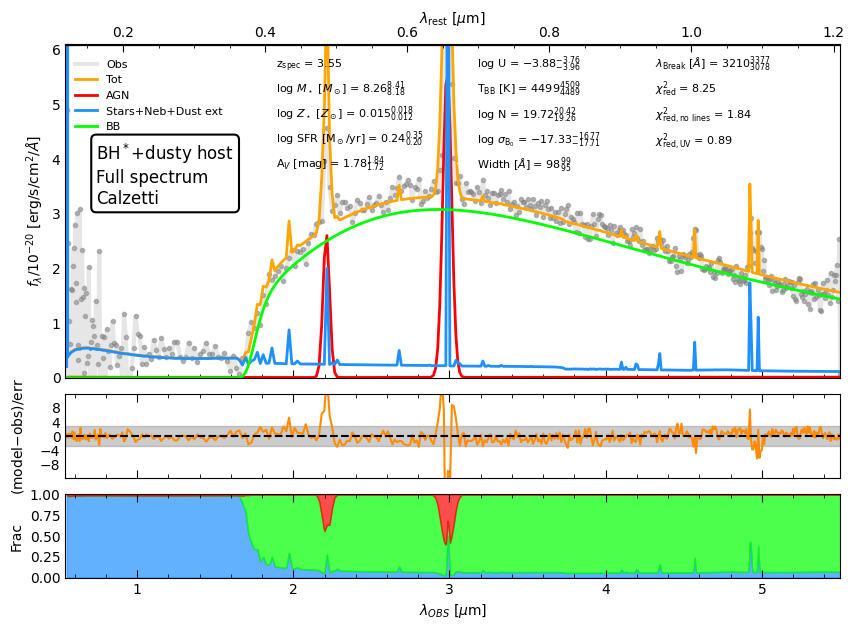}\\[-2mm]
    \includegraphics[width=\linewidth]{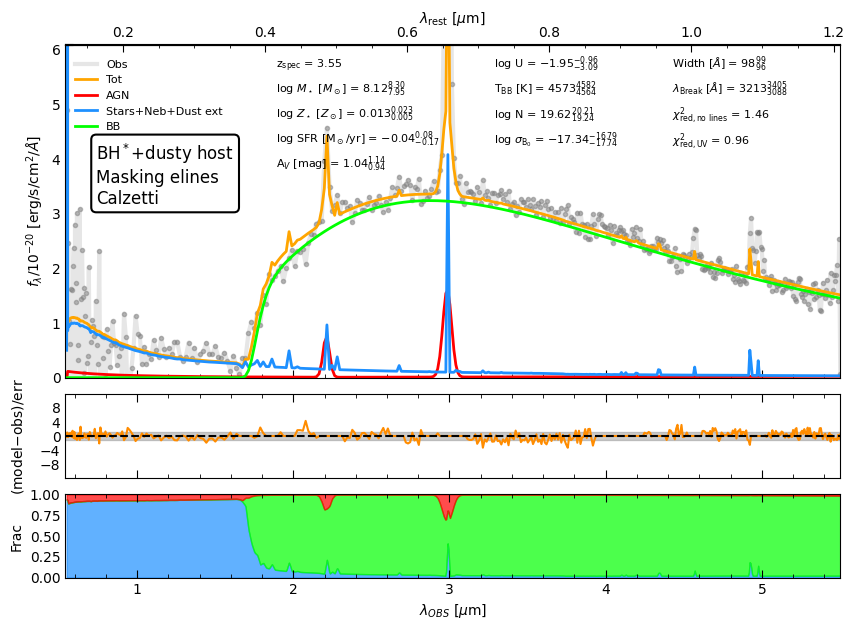}
}
    \caption{Same as Fig. \ref{fig:cliff_dusty} but using a \citet{Calzetti2000} dust attenuation law. The use of a Calzetti law yields higher $A_V$ and $M_\star$ values when the continuum is considered.}
    \label{fig:cliff_dusty_calzetti}
\end{figure*}

We present the results for the BH$^*$+dusty host using a Calzetti law in Fig.~\ref{fig:cliff_dusty_calzetti}. A summary of the derived properties can be consulted in Table~\ref{tab:properties}. When the full spectrum is considered, the results of the fit are consistent with those reported using an SMC law: a star-forming, $10^{8.3}\,M_\odot$ host with $A_V\sim1.8$~mag. However, the fit in the UV improves when we consider a Calzetti law, as this law is less steep in the UV and consequently suppresses the stellar component to a lesser extent.

If emission lines are masked, $M_\star$ drops ($\sim10^{8.1}\,M_\odot$), but this value is still higher than the results obtained for the SMC law based on the continuum alone. Metallicity is low, but higher than in the SMC case (log $Z/Z_\odot = 0.013$). Moreover, the required $A_V$ decreases only down to $\sim1$~mag, significantly beyond the limits of the BH$^*$+dust-free host $A_V$ prior. The rest of the parameters are consistent with those obtained using an SMC law, including the BB temperature and the reduced $\chi^2$ values.

This has important implications for the nature of LRD host galaxies. Depending on the assumed attenuation law and the stellar contribution to the emission lines, we infer significant dust attenuation in these systems, ranging $0.5-1.8$~mag in the case of The Cliff. For context, normal star-forming galaxies of similar $M_\star$ show $A_V\sim0.3$~mag at these $z$ \citep{Merida2023}.
However, we cannot discriminate between a moderately or highly dust-obscured host through the data, as these $A_V$ values remain consistent with the upper limits found for the LRDs' dust mass from Atacama Large Millimeter/submillimeter Array (ALMA) 1.3~mm continuum observations ($\sim10^6\,M_\odot$, \citealt{Casey2025}), assuming effective radii of $\sim100$~pc (\citealt{Baggen2023}, \citealt{Guia2024}) and the $M_d-A_V$ calibration from \citet{Ferrara2024a}.

Differences in the $A_V$ upper limits depending on the selected attenuation law were also reported in \citet{Chen2025b}. That work used the SMC, Milky Way, and Orion Nebula attenuation laws to estimate the upper limits on $A_V$ of LRDs from their predicted IR fluxes using data from the JWST Mid-Infrared Instrument, \textit{Herschel}, and ALMA. They found that extinction curves that are steeper in the UV (like SMC) result in stronger UV attenuation for a given $A_V$, leading to larger reprocessed IR fluxes, and thus a stricter $A_V$ upper limit, in line with our results. These authors derived upper limits of ${A_V \leq 1-1.5}$~mag for the brightest LRDs to date, A2744-45924 \citep{Labbe2024environment} and RUBIES-BLAGN-1 \citep{Wang2025}, and for stacked SEDs from a large sample of LRDs, further supporting a dusty nature for the hosts of these objects.

\subsection{The role of the AGN continuum in the UV}
\label{sec:discussion_agn}

Our setup allows for the possibility of leakage of the AGN continuum into the UV, even for low values of the \texttt{f5100A} parameter, which sets the normalization of the AGN component. This regime corresponds to negative values of the slope $\alpha_\lambda$, which governs the UV shape of the AGN continuum. 

We found negative $\alpha_\lambda$ values in all configurations (see also the $\alpha_\lambda$ values derived using a Calzetti dust attenuation law), except for the BH$^*$+dust-free host setup based on the continuum (see Table~\ref{tab:properties}). However, $\alpha_\lambda$ plays a negligible role in the BH$^*$+dust-free host configuration, as the \texttt{f5100A} parameter is sufficiently low to effectively suppress the AGN continuum. This is reflected by the fractional contribution of each component to the total model shown in each panel in Fig.~\ref{fig:cliff_pure}.

In contrast, a small AGN contribution is present when using the BH$^*$+dusty-host configuration (see Fig.~\ref{fig:cliff_dusty}). However, this contribution remains very low: the AGN UV component accounts for only $2-4$\% of the total best-fitting model, with the lower value obtained when fitting the full spectrum. The inferred AGN fractions---8\% of the UV model flux when using the continuum and 19\% when fitting the full spectrum---are comparable to the intrinsic degeneracies of the \texttt{Bagpipes} fitting. Nevertheless, they may also be hinting at a genuine physical contribution from the AGN, such that the UV continuum is a combination of AGN and host galaxy emission. So far, no UV emission lines have been detected in The Cliff, but the available medium‑resolution spectroscopic data do not extend blueward of H$\alpha$, preventing any direct assessment of the UV line spectrum.

In practice, AGN leakage in the model arises from the interplay between the stellar, nebular, and AGN components through dust attenuation, which does not affect the AGN model by construction. To some extent, the code attempts to suppress the stellar contribution to the UV continuum (powered by young stellar populations that produce a hard ionizing radiation field) using dust attenuation. This effectively weakens the coupling between the UV continuum and the nebular line production. At the same time, a modest AGN contribution helps reproduce the observed UV continuum without increasing the ionizing photon budget.

The preference of the model for a dustier solution with modest AGN UV leakage cannot be dismissed outright, as the resulting reduced $\chi^2$ values remain consistent with statistically acceptable fits. In other words, if such leakage were physically present, the model would still be capable of capturing it while producing a BH$^*$-like solution. 

At the same time, this leaves open the possibility that the UV continuum could be entirely AGN‑driven, with negligible contribution from the host. To test this scenario, we reran the code after removing the stellar, nebular, and dust attenuation components from the fit. The corresponding results, based on the continuum, are presented in Appendix~\ref{app1}. The overall reduced $\chi^2$ value is worse here, especially in the UV. The AGN is mostly suppressed through the normalization in the optical, but the UV slope makes it possible for the AGN to produce most of the UV emission. However, the best-fitting model underestimates the flux at $\lambda>0.2\,\mu$m, potentially ruling out a host-less scenario in this LRD.

In Appendix~\ref{app1} we also divided the $\alpha_\lambda$ prior into two distinct regimes, an unobscured AGN regime ($\alpha_\lambda = [-2.8, -1.8]$) and an obscured AGN regime ($\alpha_\lambda = [0, 1]$). This experiment effectively forces the model to explore both a quasar-like UV continuum (i.e., allowing for AGN UV leakage) and a reddened AGN solution (i.e., suppressing AGN UV leakage by construction). We found that the two regimes yield statistically indistinguishable fits. In other words, a purely stellar‑driven UV continuum and a mixed AGN+stellar UV continuum remain degenerate solutions.

It is worth noting that a UV continuum that is not solely host-driven, together with weak [OIII] emission, such as that observed in The Cliff, is a scenario that can be naturally explained within the unified AGN framework for LRDs proposed by \citet{Madau2026}. According to this theory, LRDs host a super-Eddington central engine and differ primarily from compact Type I AGNs in terms of orientation and line-of-sight obscuration. The central engine is modeled as a geometrically thick radiation-pressure–supported accretion flow whose funnel produces strongly anisotropic, intrinsically blue ionizing continua. This is coupled to an equatorially concentrated BLR and a dusty reprocessing screen with a modest covering factor. \citet{Madau2026} pointed out that weak [OIII] emission in LRDs with the strongest Balmer breaks may not require a host-dominated UV continuum, as is the case for luminous quasars.
The [OIII] equivalent width in these systems is known to decrease with increasing luminosity \citep{Baldwin1977}. This would imply high intrinsic luminosities in [OIII] weak, Balmer break LRDs, which could be the case for The Cliff, as strong Balmer breaks correspond to the most dust-attenuated sightlines. The compact Type I AGN counterparts of these sightlines potentially overlap with the quasar regime. 

\subsection{The star formation and black hole growth history of The Cliff}
\label{sec:discussion_sfh}

In this subsection we attempt to examine the SFH of The Cliff's host and relate it to the BH accretion history. For this exercise, we cannot rely on the most recent SFH, as it is inferred from the emission lines powered by star formation. At present, we cannot disentangle the AGN contribution from the nebular emission of the host, making the SFH at $<10$~Myr highly uncertain due to the likely contamination of the key SFR-tracing emission lines by the AGN. Moreover, the interplay between dust attenuation, stars, and nebular emission can also bias the SFR estimates on longer timescales ($\sim100$~Myr), which are probed by the UV continuum. These limitations imply that the results presented here should be interpreted with caution and regarded as exploratory.

Despite these caveats, there is observational evidence suggesting that at least part of the nebular emission still traces recent star formation. \citet{Asada2026} found a tight correlation between the narrow-line component and the UV-to-H$\alpha$ luminosity ratio of a sample of $z>5$ LRDs, consistent with that expected for young starburst galaxies. This trend was also observed for the broad H$\alpha$ component, albeit with a larger scatter. This implies that, in principle, the narrow H$\alpha$ component of The Cliff can be used to estimate its most recent star formation. Such an interpretation is further supported by the physics of the BH$^*$ model, as the gas envelope would prevent any AGN-driven narrow line emission from contaminating the spectrum.

The complex H$\alpha$ line profile of The Cliff was explored in \cite{deGraaf2025} using \texttt{unite} \citep{Hviding2025} and a G395M spectrum of this target. Their fiducial model assumed a broad Lorentzian and narrow Gaussian emission component, a narrow Gaussian absorption component,
and a linear continuum. Assuming that the narrow component is only powered by stars, we can use the \citet{Kennicutt1998} calibration for a \citet{Chabrier2003} IMF and estimate the SFR from H$\alpha$. Considering a flux of 1.3$\times10^{-17}$~erg/s/cm$^2$ \citep{deGraaf2025}, we estimate an SFR(H$\alpha) = 7 - 26\,M_\odot$/yr, where the lower value corresponds to $A_V=0$~mag and the upper value to $A_V \sim 1.8$~mag.

However, the full width at half maximum of the narrow H$\alpha$ component reported in \citet{deGraaf2025} is $478\pm95$~km/s. This width is typical of AGN narrow lines, much higher compared to the typical values of $100-200$~km/s of their stellar counterparts.
This strongly suggests that the narrow H$\alpha$ emission in The Cliff is not purely stellar (see \citealt{Sok2026}), and therefore cannot be reliably used to trace star formation. As a result, the most recent SFH of LRDs remains unconstrained under the current limitations. A deeper understanding of the origin of LRDs' emission lines---and an assessment of whether such features can be produced within the BH$^*$ framework---is essential before robust SFR estimates can be derived.

\begin{figure}[htp]
    \centering
    \includegraphics[width=.9\linewidth]{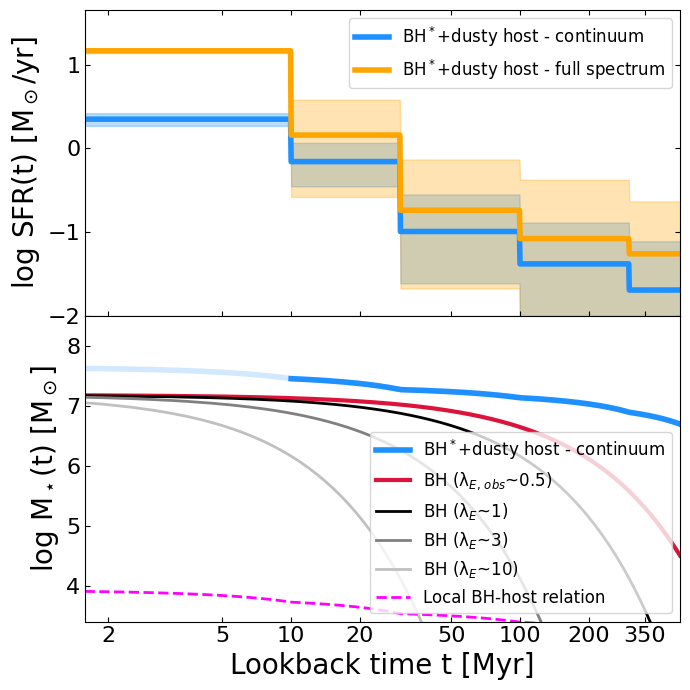}
    \caption{\textit{Top}: SFHs derived using the BH$^*$+dusty host configuration based on the full spectrum (orange) and the continuum (blue). Shaded regions denote the area between the 16th and 84th percentiles, with the median represented as a solid line. \textit{Bottom}: Mass assembly history of The Cliff according to the BH$^*$+dusty host configuration based on the continuum (blue). The color of the curve is lighter at $t<10$~Myr as a reminder of the lack of proper emission line constraints. Hypothetical growth tracks of the BH are included, assuming a final $M_{\mathrm{BH}}=10^{7.18}\,M_\odot$ and different Eddington accretion ratios. The $\lambda_E$ of $\sim 0.5$ (red line) corresponds to the observed value. We depict three additional cases of Eddington (black) and super-Eddington accretion (gray and silver). The evolution of a BH along the BH--host local relation from \citet{Reines2015} is also displayed (dashed fuchsia line).}
    \label{fig:sfh}
\end{figure}

With caveats about the reliability of SFR estimates in mind, especially at lookback times $<$10~Myr, we proceed to examine The Cliff's SFH. 
Figure~\ref{fig:sfh} shows the SFH for The Cliff based on the BH$^*$+dusty host setting, derived using the full spectrum and only the continuum. These tracks show SFR $\sim 0.01 - 1\,M_\odot$/yr at lookback times $\gtrsim 10$ Myr. If the full spectrum is considered, SFR increases to $\sim10~\,M_\odot$/yr at more recent lookback times.

This figure also includes the mass assembly history of The Cliff, derived by integrating the SFH while assuming a return fraction of 0.3 (i.e., as expected for a \citealt{Chabrier2003} IMF). We do this for the BH$^*$+dusty host configuration based on the continuum. Note that, given its relative youth, assuming a constant return fraction of $0.3$ likely overestimates the amount of recycled gas in this system. Thus, the $M_\star$(t) could be somewhat higher, by up to $\sim0.3$~dex, if we assumed lower return fractions. According to our results, the host galaxy increased its $M_\star$ by only $\sim0.4$~dex in the period ranging $\sim10-300$~Myr in lookback time.

The theoretical predictions of \citet{Inayoshic} describe the assembling of the LRD hosts through an ${\mathrm{SFR}\geq1\,M_\odot\mathrm{yr}^{-1}}$ starburst, which led to the formation of a ${\sim10^7\,M_\odot}$ compact (${\leq100}$~pc) stellar cluster within $\sim10$~Myr. In contrast, our results point to a much smoother evolutionary path: the galaxy appears to have been building up its $M_\star$ over a longer period, with a $10^7\,M_\odot$ host already in place $\sim200$~Myr.

Even though we do not have information about the coevolution of The Cliff's host and the central engine, we can study the BH mass assembly history under certain assumptions. We can consider mass growth via accretion onto less massive seeds \citep{Madau2014} and continuous gas supply along with uncorrelated accretion events. This ensures a soft duty cycle of $\sim100$\%, which is a fair assumption for a gas-rich galaxy at these early epochs \citep{Zubovas2021}. Following \cite{Madau2014} and if BH accretion is proportional to the BH mass, $\dot{M}_{BH}\propto M_{BH}$, as given by the $M-L$ relation, we can derive the time-dependent mass of the central BH as
\begin{equation}
\begin{aligned}
    \frac{dM_{BH}}{dt} = \frac{(1-\epsilon)}{\epsilon}\frac{L}{c^2} \; \mathrm{with}\\
    M_{BH}(t) = M_{BH, 0}\,exp \left( \frac{(1-\epsilon)}{\epsilon} \frac{\lambda_E}{t_E} (t-t_0)\right),
\end{aligned}
\end{equation}
where $M_{BH,0}$ is the mass of the BH seed, $\epsilon$ is the accretion efficiency, $\lambda_E$ is the Eddington accretion ratio, and $t_E$ is the Eddington timescale, defined as ${t_E=M_{BH}c^2/L_E=0.44\mu_e^{-1}\,\mathrm{Gyr}}$ \citep{Madau2014}. $\mu_e$ is the mean molecular weight per electron, which is 1.15 for primordial gas. Assuming this $\mu_e$ value and setting $\epsilon=0.1$, we computed $M_{BH}(t)$ backward in time, starting with the current estimated $M_{BH}$.

The estimated $M_{BH}$ for this object (without including a dust correction) is log ($M_{BH}/M_\odot$) = $7.18_{-0.06}^{0.07}$ \citep{deGraaf2025}, and thus $M_{BH}/M_\star = 0.26$ (assuming a $M_\star=10^{7.7}\,M_\odot$), well above the $M_{BH} - M_\star$ local relation \citep{Reines2015}, as usually found for LRDs. This implies that The Cliff is a BH-dominated system. 
Its bolometric luminosity, assuming the scaling relations between $L_{\mathrm{H}\alpha}$, $L_{5100}$, and $L_{\mathrm{bol}}$ of \citet{Greene2005} and \citet{Shen2020}, is ${\sim1\times10^{45}}$~ erg~s$^{-1}$ \citep{deGraaf2025}, which translates into a BH accretion rate of 0.18~$M_\odot$/yr.
The Eddington accretion ratio $\lambda_E$ of The Cliff, defined as the ratio between the bolometric and Eddington luminosities, is $\sim0.5$. This object is thus accreting in a sub-Eddington regime. However, note that this approach assumes that the AGN is the primary driver of the H$\alpha$ emission, whereas the origin of the LRDs' broad Balmer lines is still a matter of debate. Additionally, $M_{BH}$ should not exceed a few times $10^7\,M_\odot$ according to some BH$^*$ theoretical works (see $M_{\mathrm{crit,\,LRD}}$ in \citealt{Inayoshic}), which is the mass at which the BH feeding rate is supposed to fall below the Eddington limit, preceding the end of the LRD phase.
However, most reported LRD $M_{BH}$ values are larger than this value (see \citealt{Juodzbalis2025_b} and references therein), and the measured ${\lambda_E}$ is already sub-Eddington in The Cliff. This points to a potential issue in the BH$^*$ model or a lack of accurate estimates of the BH properties in LRDs.

We estimated the mass assembly history of the BH for a constant  ${\lambda_E=0.5}$ and three additional cases, corresponding to Eddington and super-Eddington accretion (see Fig.~\ref{fig:sfh}).
All these growth tracks yield a BH growth that is incompatible with the local BH--host scaling relation, which predicts a final $M_{BH}$ for The Cliff of $\sim10^4\,M_\odot$ based on the host's stellar mass. Such a low $M_{BH}$ value is not unreasonable, as \cite{Liu2026} estimated a BH plus gas mass of $\sim10^4\,M_\odot$ for the local LRD analog, ``The Egg". Similarly, \cite{Sun2026b} obtained $M_{BH}$ of $\sim10^{4-5}\,M_\odot$ from stellar-atmosphere fits to host-subtracted stacks built from 117 LRDs, with $L_{\mathrm{bol}}/L_{E}$ ratios reaching $\sim$50.

However, the current offset between the $M_{\rm BH}$ inferred from H$\alpha$ and that predicted by the BH--host scaling relation does not necessarily imply that the BH and its host have evolved off the local relation throughout their entire histories. Short episodes of super-Eddington accretion (e.g., \citealt{Li2023}, \citealt{Lupi2024}, \citealt{Trinca2024}, \citealt{Quadri2025}) could instead drive rapid BH growth at late times, moving the system away from the relation.

An alternative interpretation is provided by the direct collapse BH scenario (e.g., \citealt{Cenci2025}, \citealt{Baggen2026}, \citealt{Pacucci2026}). In this framework, the tension in the host--BH coevolution naturally disappears, as the BH forms and grows independently of the stellar component. Subsequent interactions, such as migration of the BH into a low-mass companion galaxy, whose UV radiation suppressed molecular hydrogen cooling, thereby allowing the formation of these direct collapse BHs, could then explain the large $M_{BH}/M_\star$ ratio estimated in The Cliff.

\section{Conclusions}

In this work, we tested the assumptions of the BH$^*$ model by performing a full UV-to-optical spectral fitting of The Cliff \citep{deGraaf2025} using JWST/NIRSpec PRISM data and a modified version of \texttt{Bagpipes} that includes BB emission and gas absorption at the Balmer limit.
Our modeling naturally generates the BH$^*$ picture in this LRD without requiring strong constraints, indicating that this solution is driven by the data rather than imposed by the priors, and thereby providing a robust test of the BH$^*$ scenario. 

Furthermore, according to our results, there is a low-mass galaxy host in The Cliff, rendering the host-less scenario unlikely for this and similar LRDs. Our method thus provides a useful avenue for characterizing the hosts of LRDs even when they cannot be spatially resolved in imaging. The stellar component required by the fits constitutes potential evidence of the presence of a host galaxy in LRDs.

According to our fits to the spectral continuum, The Cliff's host is a $\sim10^{7.7}\,M_\odot$\ metal-poor galaxy that evolves along the star-forming MS, in line with the expectations for LRD hosts within the BH$^*$ framework. However, this host is not dust-free, as it is subject to non-negligible dust attenuation (${A_V\sim0.5}$~mag under the SMC dust law) acting on the stellar and nebular components. Moreover, adopting a \citet{Calzetti2000} attenuation law instead of a SMC law yields larger $M_\star$ and $A_V$ values ($\sim10^8\,M_\odot$ and $A_V\sim1$~mag), which suggests an even more heavily obscured host.

In terms of the envelope, which dominates the optical continuum, we infer a BB temperature of $\sim4570$~K. This value is fully consistent with theoretical expectations within the BH$^*$ framework, where the envelope behaves as a photosphere near the Hayashi limit.

In the BH$^*$ scenario, the AGN emission is reddened by this envelope and thus does not contribute to the continuum. However, AGN continuum leakage in the UV is allowed and sometimes preferred by the code, which still preserves the overall BH$^*$ picture, even with a small contribution of the AGN to the UV continuum. 
A quasar-like UV continuum AGN (i.e., stellar+AGN-powered UV) is statistically indistinguishable from that of an obscured AGN (i.e., stellar-driven UV).
A better understanding of the gas envelope's clumpiness, as well as the origin of potential high-ionization UV lines, will be key to shedding light on the AGN contribution to the UV emission of LRDs.

Finally, our fits also provide tentative insight into the evolutionary history of The Cliff. Its SFH indicates a smooth evolutionary path, with a $M_\star$ of $10^7\,M_\odot$ already in place 200~Myr ago. The BH–host evolution consistently lies above the expectations from typical BH–host scaling relations, especially at recent times.
Theories invoking direct collapse BHs \citep{Pacucci2026} or the unified AGN framework \citep{Madau2026} offer potential explanations for the potential lack of BH--host coevolution and the nature of the UV emission in LRDs that are not host-related. However, further testing with larger statistical samples of LRDs and higher-resolution spectra is required to robustly test these models, as well as the BH$^*$ scenario.

\begin{acknowledgements}

We thank Jenny E. Greene, David Setton, and Yilun Ma for valuable discussions and feedback that improved this work.\\
This research was enabled by Canadian Space Agency grants 18JWST-GTO1 and 24JWGO3A12 and Natural Sciences and Engineering Research Council (NSERC) of Canada grants RGPIN-2020-06023 and RGPAS-2020-00065.\\
The data products presented herein were retrieved from the Dawn JWST Archive (DJA). DJA is an initiative of the Cosmic Dawn Center (DAWN), which is funded by the Danish National Research Foundation under grant DNRF140.\\
This research used the Canadian Advanced Network For Astronomy Research (CANFAR) platform operated in partnership by the Canadian Astronomy Data Centre and The Digital Research Alliance of Canada with support from the National Research Council of Canada, the
Canadian Space Agency, CANARIE, and the Canada Foundation for Innovation.\\
This work is based on observations made with the NASA/ESA/CSA \textit{James Webb} Space Telescope, associated with the program GO-4233. The data were obtained from the Mikulski Archive for Space Telescopes at the Space Telescope Science Institute, which is operated by the Association of Universities for Research in Astronomy, Inc., under NASA contract NAS 5-03127 for JWST. 
\end{acknowledgements}

\bibliographystyle{aa}
\bibliography{aa60857-26}

\begin{appendix}

\onecolumn

\section{The effect of the absorption component}
\label{app2}

\begin{figure}[htp]
    \centering
        \sidecaption
 \includegraphics[width=0.7\linewidth]{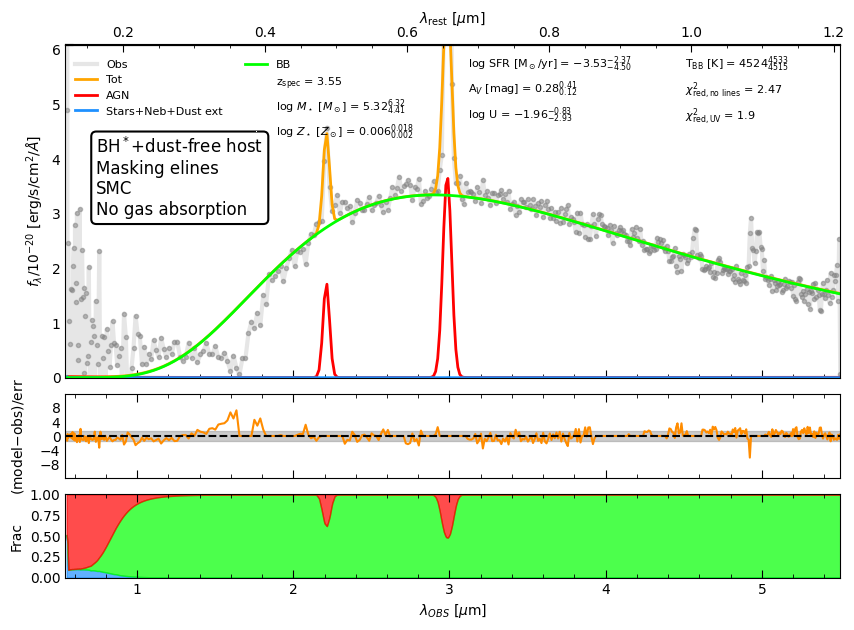}
    \caption{Best-fitting model for The Cliff using the BH$^*$+dust-free host configuration and an SMC law while masking the emission lines and not including the absorption component. See Fig.~\ref{fig:cliff_pure} for a full description of the markers and color codes. The BB model and the total model overlap in most of the wavelength range. Without this feature, the model overestimates the flux near the Balmer break.}
    \label{fig:abs}
\end{figure}

Figure~\ref{fig:abs} shows a fit to The Cliff using the BH$^*$+dust-free host configuration while masking the emission lines and not including the gas absorption component (see Sect.~\ref{sec:method} for a detailed description of this feature). 
The BB model does not fit the data in the vicinity of the break, overestimating the flux and producing an excess clearly reflected in the residuals. \texttt{Bagpipes} attributes most of the continuum flux to the BB, eliminating the need for a stellar component. Although not plotted here, the same outcome is observed in the BH$^*$+dusty host configuration. 

\section{An AGN nature for the UV continuum?}
\label{app1}
\begin{figure}
    \centering
        \sidecaption
    \includegraphics[width=0.7\linewidth]{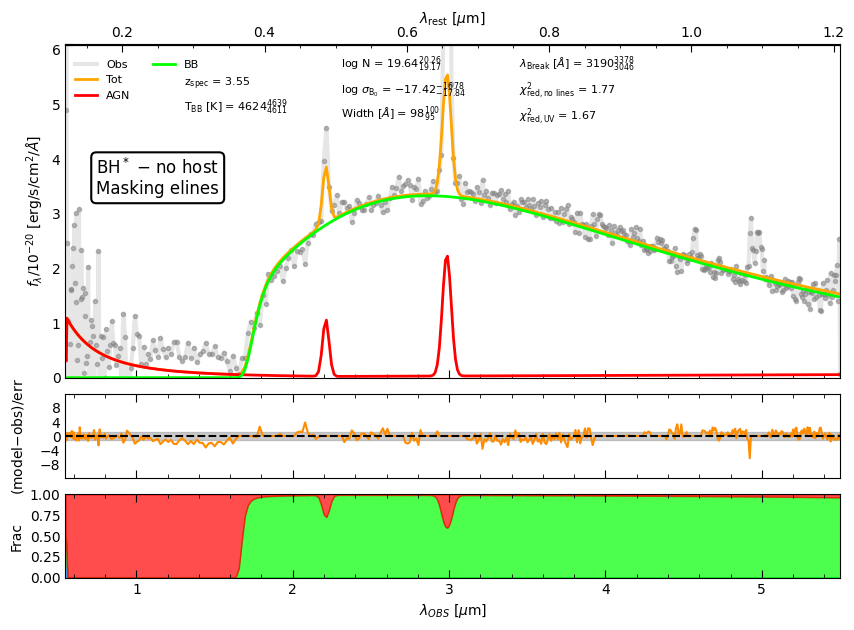}
    \caption{Best-fitting model for The Cliff masking the emission lines and excluding the stellar, nebular, and dust attenuation components from the fit. See Fig.~\ref{fig:cliff_pure} for a full description of the markers and color codes. The AGN model and the total model overlap in the UV. Without the host galaxy, the model underestimates the flux at rest-frame ${\lambda>0.2\,\mu}$m.}
    \label{fig:no_host}
\end{figure}

\subsection{A host-less scenario}

Our model includes stellar, nebular, dust attenuation, AGN, BB, and gas absorption components by construction. Our fits favor a scenario whereby the UV continuum is mostly driven by the host galaxy. In this first section of the appendix, we refitted The Cliff's continuum allowing only an AGN and BB + absorption components. The results are displayed in Fig.~\ref{fig:no_host}.

The optical continuum is driven by the BB, whereas the UV is powered by the AGN. However, this model underestimates the flux at rest-frame $\lambda>0.2\,\mu$m, suggesting the need for an additional component.

\begin{figure*}[!htp]
    \centering
        \sidecaption
        \parbox{0.70\textwidth}{
    \includegraphics[width=\linewidth]{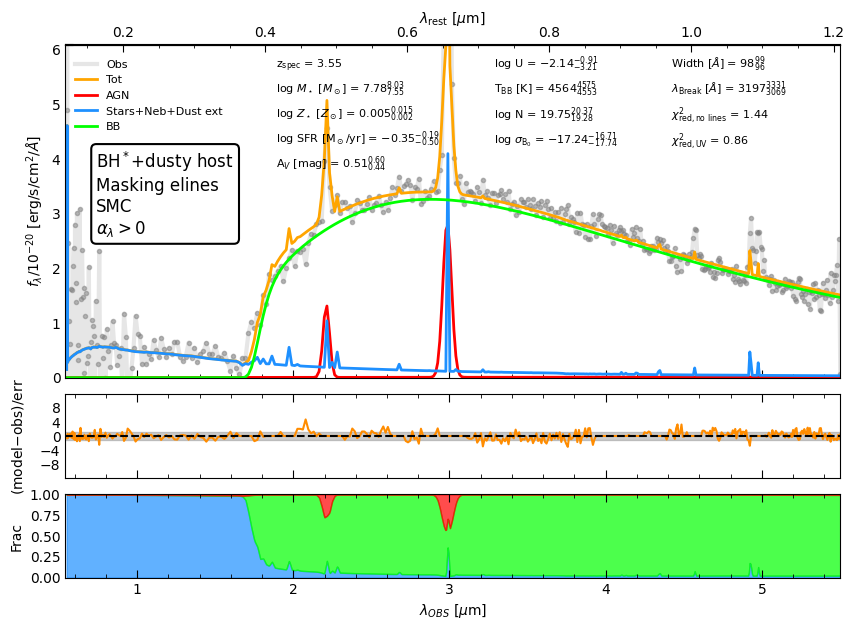}\\[-2mm]
    \includegraphics[width=\linewidth]{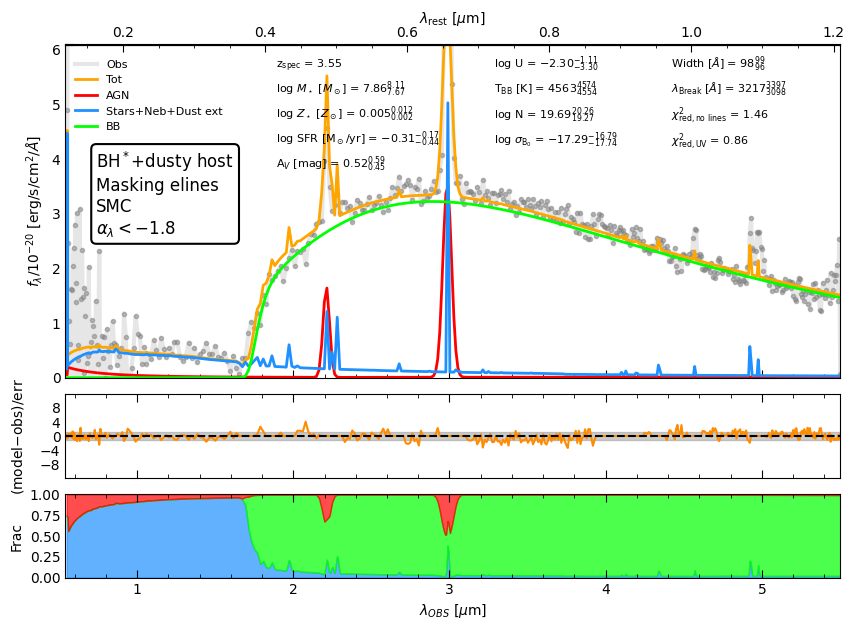}
}
    \caption{Best-fitting models from \texttt{Bagpipes} using the BH$^*$+dusty host configurations and the continuum emission of The Cliff, imposing an obscured AGN (\textit{top}) and an unobscured AGN (\textit{bottom}) model. The two are statistically indistinguishable. See Fig.~\ref{fig:cliff_pure} for a full description of the markers and color codes.}
    \label{fig:cliff_dusty_agn}
\end{figure*}

\subsection{The effect of the AGN $\alpha_\lambda$ slope}

In this work, we assumed an AGN model in which the emission from the accretion disk can potentially arise in the UV even with a low AGN normalization factor.  
In this second section of the appendix, we explore different regimes by narrowing down the $\alpha_\lambda$ prior, leaving the rest of the priors untouched. We forced an obscured solution by setting $\alpha_\lambda=[0,\,2]$, and an unobscured solution by setting $\alpha_\lambda=[-2.8,\,-1.8]$. In this way, we are changing the effective visibility of the accretion disk. We used the BH$^*$+dusty host configuration and fitted only the continuum, masking the lines.

The best-fitting models from \texttt{Bagpipes} obtained with these narrower $\alpha_\lambda$ priors are displayed in Fig.~\ref{fig:cliff_dusty_agn}. The parameters from the stellar + nebular components and BB + absorption are consistent with those listed in Table~\ref{tab:properties}. The obscured AGN model yields no AGN contribution to the UV, while the unobscured AGN model leads to some AGN leakage in the UV. These models are statistically indistinguishable, as reflected by their $\chi^2$ values.

\end{appendix}
\end{document}